\documentstyle[12pt]{article}
\newcommand{\thetabar}{\overline{\theta}}

\newcommand{\sigmabar}{\overline{\sigma}}
\newcommand{\Sigmabar}{\overline{\Sigma}}
\newcommand{\lambdabar}{\overline{\lambda}}
\newcommand{\Phibar}{\overline{\Phi}}
\newcommand{\phibar}{\overline{\phi}}
\newcommand{\psibar}{\overline{\psi}}
\newcommand{\Pibar}{\overline{\Pi}}
\newcommand{\pibar}{\overline{\pi}}
\newcommand{\h}{\hspace{2mm}}
\newcommand{\Fbar}{\overline F}
\newcommand{\alphadot}{\dot{\alpha}}
\newcommand{\gbar}{\overline g}

\catcode`\@=11

\begin{document}
\baselineskip .25in
\renewcommand{\thefootnote}{\dag}
\newcommand{\numero}{SHEP 93/94-12}   %Enter SHEP preprint number

\newcommand{\titre}{A SUPERSYMMETRIC EFFECTIVE CHIRAL
LAGRANGIAN }
\newcommand{\auteura}{K.J. Barnes}
\newcommand{\auteurb}{D.A. Ross}
\newcommand{\auteurc}{R.D. Simmons }
\newcommand{\place}{Department of Physics,\\University of
Southampton\\
Southampton SO9 5NH \\ U.K. }
\newcommand{\beq}{\begin{equation}}
\newcommand{\eeq}{\end{equation}}

\newcommand{\abstrait}{  We construct in a manifestly
supersymmetric form
the leading and subleading terms in momentum for an effective
supersymmetric
chiral Lagrangian in terms of complex pions and their
superpartners. A
soft supersymmetry breaking term is included and below the
supersymmetry
breaking scale
the Lagrangian reduces to the usual bosonic chiral Lagrangian
in terms of
real pions.}
\begin{titlepage}
\hfill \numero  \\

\vspace{.5in}
\begin{center}
{\large{\bf \titre }}
\bigskip \\ by \bigskip \\ \auteura \bigskip \\  \auteurb \\
  \bigskip  and \bigskip \\ \auteurc
    \bigskip \\ \place \bigskip \\

\renewcommand{\thefootnote}{\dag }
\vspace{.9 in}
{\bf Abstract}
\end{center}
\abstrait
 \bigskip \\
\end{titlepage}

Supersymmetry is an attractive candidate for constructing
models beyond the Standard
Model \cite{susy1} mainly because of the cancellation of
logarithmic divergences in
higher order corrections thereby providing a solution to the
naturalness
nature of the hierarchy problem which inevitably occurs when
attempts
are made to construct models in which new physics is
postulated between
currently accessible energies and the Planck scale. However
such
supersymmetric extensions of the Standard Model still require
the existence of
fundamental Higgs particles, albeit in chiral supermultiplets.
There has
to date been no evidence of the existence of fundamental
scalars and one
is naturally led to speculate that such Higgs particles are
not fundamental particles, but are low energy manifestations
of the effects
of some new physics at a higher scale, which forms condensates
when the
couplings associated with that new physics become strong.

It is interesting to consider the possibility that such new
physics exists
in conjunction with supersymmetry and that furthermore the
condensation scale
of this new physics is large compared with the supersymmetry
breaking scale.
It has been pointed out \cite{niles} that the conventional
fermion condensate
of standard technicolour theories cannot occur without at the
same time
breaking supersymmetry. Such a breaking would not be
consistent with the
scenario proposed here since the breaking scales would then
necessarily be the same.
However as pointed out in ref.\cite{niles}, if it were not the
fermions
but a composite formed out of the scalar supermultiplets of
the fermions
 that acquired a vacuum expectation
value, then it is indeed possible to break the internal chiral
symmetry,
whilst at the same time preserving the supersymmetry. If that
were actually
the case then there would be
an important energy regime between the supersymmetry  breaking
scale
and the chiral symmetry breaking scale where physical
processes could
be described in terms of a supersymmetric extension of the
effective
chiral Lagrangian proposed by Gasser and Leutwyler \cite{gl}.
In such a model the pions would be replaced by chiral
supermultiplets with
each pion having an associated Majorana fermion (the
``piino'').
The effective action as in the case of ref.\cite{gl}
has an infinite
number of terms with arbitrary coefficients, but can be
systematically
expanded in powers of momentum (scaled by the chiral-symmetry
breaking scale). Internal loops of the ``piinos'' would have a
significant
effect on
 calculations which
have been hitherto performed using the chiral Lagrangian
technique and there
would be particularly interesting consequences for the
restoration of
 unitarity as more  terms in the effective Lagrangian
are taken into
consideration.

The requirement that  the effective Lagrangian must be
invariant under
 (N=1) supersymmetry transformations order by order in
 momentum severely restricts the terms that
it may contain. As a method of obtaining the leading  term, we
begin
by considering a supersymmetric Higgs model. It is important
to note
that such a model differs in a crucial way from the effective
chiral
Lagrangian model proposed in that it contains fundamental
Higgs fields
other than the Goldstone bosons and their superpartners. We
use this model for two purposes. The first is
as a demonstration  of how a model with a linearly
realised
chiral symmetry and supersymmetry can be broken spontaneously
into
a model in which only the vector symmetry is linearly
realised, and the
axial symmetry is realised by Goldstone bosons, which
transform in a non-linear
 manner and at the same time preserve the supersymmetry.
The second is as an aid
to construct the leading momentum term of the effective
theory.

The most general supersymmetric action of $n$ chiral
superfields, $\Phi_i$, may be written as
\begin{equation}
I = \int d^{8}z\Phibar_{i}\Phi_{i} + \int d^{6}sW(\Phi) + \int
d^{6}\overline{s}\overline{W}(\Phibar)
\end{equation}
where the superpotential $W(\Phi)$ is a functional of chiral
superfields
 alone. Because we wish to construct the smallest chiral symmetry (for
simplicity)
we choose the manifestly SU(2)$_L \otimes$SU(2)$_R$ invariant
superpotential
\begin{equation}
W(\Phi) = (\Sigma^{2} + \Pi_{a}\Pi_{a} - f^{2}_{\pi})\Phi
\end{equation}
where in components we have
($y^m=x^m+i\theta\sigma^m\overline{\theta}$)
\begin{eqnarray}
\Sigma (x,\theta,\thetabar)&=&\sigma(y) +
\sqrt{2}\theta\lambda_{\Sigma}(y) +
\theta^{2}F_{\Sigma}(y)\nonumber\\
\Pi^{a} (x,\theta,\thetabar)&=&\pi^{a}(y) +
\sqrt{2}\theta\lambda^{a}(y) +
\theta^{2}F^{a}(y)\nonumber\\
\Phi (x,\theta,\thetabar)&=&\phi(y) +
\sqrt{2}\theta\lambda_{\phi}(y) +
\theta^{2}F_{\Phi}(y)\
\end{eqnarray}
where  $\sigma^m=(-1,\tau^a)$, and $\tau^a$  are the $2\times 2$
Pauli matrices,
 \ ($a=1,2,3$).  The chiral superfields
$\Sigma$ and $\Pi^a$ transform as a $(2,\bar{2})$ under
 SU(2)$_L \otimes$SU(2)$_R$ whereas the chiral superfield
 $\Phi$ is a singlet.
Combining the $\Sigma$ and $\Pi^{a}$
fields into the matrix $H=\Sigma + i\tau . \Pi$, such that
under  SU(2)$_L \otimes$SU(2)$_R$ $H$ transforms as
$$ \ H \ \rightarrow L H R^{\dagger}, $$
our starting action becomes
\begin{equation}
I = \int d^{8}z \left[ \frac{1}{2} tr\left( H \overline{H} \right)
 + \Phi\overline{\Phi} \right]
  + \h\alpha\int d^{6}s(detH - f^{2}_{\pi})\Phi\h +
\h\alpha\int d^{6}\overline{s}(det\overline{H} -
f^{2}_{\pi})\overline{\Phi}
 \label{action1} \end{equation}
\noindent which has the potential
\[
V = F_{\Sigma}\overline{F}_{\Sigma} +
F^{a}\overline{F}^{a} +
F_{\Phi}\overline{F}_{\Phi}
\]
\begin{equation}
= \sigma\phi\sigmabar\phibar + \pi^{a}\phi
\pibar^{a}\phibar + \frac{1}{4}(\sigma^{2} + \pi^{a}\pi^{a} -
f^{2}_{\pi})(
\sigmabar^{2} + \pibar^{a}\pibar^{a} - f^{2}_{\pi})
\end{equation}
The minimum of this potential is clearly $V=0$ which may be
achieved by giving
the fields the following SU(2)$_{L}\otimes$SU(2)$_{R}$ symmetry
breaking
vacuum
expectation values (VEVs)
\footnote{ We require that $f_\pi$ be taken to
be real in order
for the model to reduce to the usual bosonic chiral model below the
supersymmetry
breaking scale.}
\begin{equation}
<\sigma>\h =\h f_{\pi} \hspace{.5in} <\pi_{a}>\h =\h <\phi>\h =\h 0
\end{equation}
Importantly, as implied by the $V=0$ minimum, no auxiliary
field acquires
a VEV with the above assignments so supersymmetry is
manifestly not
broken in this model.
Using these VEVs we proceed to evaluate the fermion and boson
mass
matrices and arrive at the following particle spectrum:
\renewcommand{\thefootnote}{\ddag}
\footnote{ The
 complex scalar fields are actually linear superpositions of
scalar and pseudoscalar fields, such that the real parts of the
massive fields transform as scalars whereas the imaginary parts
transform as pseudoscalars - for the massless fields it is the other
way around.}

\begin{itemize}
\item 3 massless complex scalars, $\pi^a$
\item 3 massless Majorana fermions, $\lambda^a$
\item 2 massive complex scalars, $\sigma$ and $\phi$
\item 1 massive Dirac fermion composed out of $\lambda_\Sigma$
and
 $\lambda_\Phi$.
\end{itemize}
where all massive particles have mass $m=2\alpha f_\pi$.

It may be pertinent to view this spectrum in the light of the
symmetries of the system. Ignoring supersymmetry for the moment,
it is clear that the kinetic terms in eq.(\ref{action1}) have an
O(8) invariance. Obviously the other two terms in this action are
designed to have  SU(2)$_L \otimes$SU(2)$_R$ invariance, and to allow
invariant couplings to chiral matter fields (although this latter
point
is not followed up in this paper). The symmetry of the action is
therefore
at least chiral  SU(2)$_L \otimes$SU(2)$_R$, but since the fields in
$H$
are complex one may ask if this can be extended. By allowing the
group
parameters in $L$ and $R$ to be considered complex, and by adjusting
the
transformation law to
$$ H \  \rightarrow \ L^{\dagger -1}HR^{-1} $$
permits the extension to transformations under  SL(2C)$ \otimes
$SL(2C) and
 coupling to corresponding matter fields. In this scenario, the real
part of
the $\sigma$ field mixes with the imaginary parts of the pions to
form a
four dimensional scalar multiplet of SL(2C) (and correspondingly the
imaginary
 part of $\sigma$ mixes with the real part of the pions to form a
pseudoscalar
multiplet). One chiral  SU(2)$_L \otimes$SU(2)$_R$ multiplet contains
the real
part of the $\sigma$ field with the real parts of the pions as usual,
and the other
 contains the complementary components. The transformations under
SL(2C)$ \otimes $SL(2C) make the eight component multiplet
irreducible.
It is immediately clear that the $detH$ remains unchanged as before.
However the
kinetic terms (notwithstanding their O(8) invariance) are only
invariant under
the chiral SU(2)$ \otimes $SU(2) subgroup of SL(2C)$ \otimes $SL(2C).
It seems,
therefore, that although the existence of the massless triplet of
pions follows
as usual from the Goldstone theorem applied to the breaking of SU(2)$
\otimes $SU(2)
down to the vector subgroup, there is no corresponding broken
symmetry of which the
massless scalar triplet are the Goldstone bosons. Instead this is a
consequence
of the supersymmetry forcing the chiral superfield to be complex.
Indeed we shall see
shortly  how soft supersymmetry breaking which still preserves the
chiral
SU(2)$ \otimes $SU(2) symmetry gives explicit masses to these
scalars.
In the absence of such soft terms one can demonstrate that the
 masslessness of the scalars is protected by supersymmetry
with a simple  calculation of the one loop correction to the scalar
and pseudoscalar masses. For the pseudoscalars, which are genuine
Goldstone bosons,
the contributions to the mass corrections from internal boson loops
and
internal fermion loops vanish separately, whereas for the scalars the
correction
vanishes by virtue of a cancellation between internal boson loops and
internal fermion
loops.

Restricting ourselves to low momenta (i.e. integrating out the
massive
degrees of freedom) we are left with three massless chiral
multiplets which
contain precisely those particles which we would expect to
appear in the
natural supersymmetric extension of the standard bosonic
chiral Lagrangian.
It is important to note here that the massless and massive
particles still
combine into supermultiplets, so that we can consistently
integrate out all
 the massive fields without violating the supersymmetry.

We may eliminate the massive degrees of freedom in a
consistent fashion
by taking the formal limit where $\alpha \rightarrow
\infty$. When this is
done we are left with the action
\begin{equation}
I = \int d^{8}z \left[\Sigma\Sigmabar + \Pi_{a}\Pibar_{a} +
\Phi\Phibar\right]
\end{equation}
with the superfields subject to the constraint
\begin{equation}
\Sigma^{2} + \Pi_{a}\Pi_{a} = f^{2}_\pi
\end{equation}
This superfield equation contains three component field
constraints
\begin{equation}
\sigma^{2} + \pi_{a}\pi_{a} = f^{2}_{\pi} \label{vev}
\end{equation}
\begin{equation}
\sigma\lambda^{\alpha}_{\Sigma} + \pi_{a}\lambda^{\alpha}_{a}
= 0
\end{equation}
\begin{equation}
F_{\Sigma}\sigma + F_{a}\pi_{a} = \frac{1}{2}(\lambda_{\Sigma}
\lambda_{\Sigma} + \lambda_{a}\lambda_{a})
\end{equation}
The first consequence of these constraints is that the
superfield $\Phi$
takes no part in the interactions - it is therefore a
spectator field
and will be ignored from now on.
Eliminating $\sigma, \lambda_{\Sigma}$, and $F_{\Sigma}$, i.e.
inserting the
above constraints into the kinetic part of the Lagrangian in
order to
obtain the leading term in the low momentum expansion, the
component field
Lagrangian becomes
\begin{eqnarray}
{\cal L}&=& g_{a \overline{b}} \left[
    - \partial_{m} \pi^{a} \partial^{m} \pibar^{b}
 +\frac{i}{2}{\cal D}\lambdabar^{b}.\sigma\lambda^{a} -
 \frac{i}{2}\lambdabar^{b}\sigma.{\cal D}\lambda^{a}\right]\nonumber\\
&+&\frac{1}{4}\lambda^{a}\lambda^{b}\lambdabar^{c}\lambdabar^{d}
 g_{a\overline d,b\overline c} -
 \frac{1}{2}\lambda^{c}\lambda^{d}
 g_{c\overline{a},d}\Fbar^{a} - \frac{1}{2}\lambdabar^{c}
\lambdabar^{d}g_{\overline{c}a,\overline d}F^{a}
+F^{a}\Fbar^{b}g_{a\overline{b}}.
\end{eqnarray}
The $\pi^a \ (\overline{\pi}^{a})$  fields are the holomorphic
(antiholomorphic)
coordinates on a K\"ahler manifold \cite{rocek} whose metric,
$g_{a\overline b}$, is given by
\begin{equation}
g_{a\overline{b}}=\delta^{ab}+\frac{\pi^{a}\pibar^{b}}{\sqrt{f^{2}
_{\pi}- \pi^{c}\pi^{c}}\sqrt{f^{2}_{\pi}-\pibar^{c}\pibar^{c}}}
\label{metric} \end{equation}
and ${\cal D}_{m}$ the covariant derivative
\begin{equation}
{\cal D}_{m}\lambda^{b}=\partial_{m}\lambda^{b}+{\Gamma}^{b}_{rs}
(\partial_{m}\pi^{s})\lambda^{r}
\end{equation}
where ${\Gamma}^{b}_{rs}$ is the connection on the K\"ahler
manifold.
The auxiliary fields are eliminated via their equations of
motion
\begin{equation}
\Fbar^{b}g_{a\overline{b}}=\frac{1}{2}\lambdabar^{c}\lambdabar^{d}
g_{\overline{c}a,\overline{d}}\hspace{1in}F^{a}g_{\overline{a}b}
=\frac{1}{2}\lambda^{c}\lambda^{d}g_{c\overline{a},d}
\label{Fterms}
\end{equation}
The inverse of the metric exists allowing us to write
\begin{equation}
{\cal L}=g_{a\overline{b}}\left[-
\partial_{m}\pi^{a}\partial^{m}\pibar^{b}
+\frac{i}{2} {\cal D}\lambdabar^{b}.\sigma\lambda^{a}
-\frac{i}{2} \lambdabar^{b}\sigma.{\cal D}\lambda^{a}\right]
+\frac{1}{4}
R_{\overline{d}a\overline{c}b}\lambda^{a}\lambda^{b}
\lambdabar^{c}\lambdabar^{d}   \label{sigma}
\end{equation}
where
$R_{\overline{d}a\overline{c}b}=g_{a\overline{d},b\overline{c}}
-g^{\overline{r}m}g_{a\overline{r},b}g_{\overline{d}m,\overline{c}}$
is the Riemann curvature tensor for a K\"ahler manifold. This
last term is
precisely the  quartic term in the fermion fields required to
complete the supersymmetry in the supersymmetric non-linear sigma
model. It
is interesting
to note how it occurs here, by elimination of the auxiliary
field in the
low momentum approximation to which we are working.

\noindent It is worthwhile pointing out at this stage that
because of the
 normalisation of the fermions relative to that of bosons, the
fermions
must be considered to have associated with them a factor of
$\sqrt{p}$,
where $p$ is the momentum scale. With this association all the
terms
in eq.(\ref{sigma}) are of order $p^2$.

\noindent This Lagrangian is supersymmetric by construction
and reduces
to the standard bosonic chiral Lagrangian when fermionic
fields are
suppressed and the scalar fields are taken to be real
- it therefore represents the first order term of a
momentum expansion of the supersymmetric chiral Lagrangian.
Following ref.\cite{zumino}  we introduce the K\"ahler
potential, $V$, defined by
\begin{equation}
g_{a\overline{b}}=\frac{\partial ^{2}V(\pi,\pibar)}
{\partial\pi^{a}\partial\pibar^{b}}
\end{equation}
allowing our action to be re-written with the superspace
Lagrangian
density $V(\Pi,\Pibar)$ - i.e. the bosonic K\"ahler potential
with the
complex scalars replaced by their chiral superfields.
Using eq.(\ref{metric}) we see that the bosonic K\"ahler
potential in this case is
\begin{equation}
V=\pi^{c}\pibar^{c}+\sqrt{f^{2}_{\pi}-\pi.\pi}\sqrt{f^{2}_{\pi
}-\pibar.\pibar}
\end{equation}

We can make contact with the usual bosonic chiral Lagrangian
by performing a change
 of variables
\begin{equation}
\Pi^a\rightarrow\Pi^a=\Pi^{'a}\frac{f_{\pi}}{\sqrt{\Pi^{'}.\Pi
^{'}}}
\sin\left(\frac{\sqrt{\Pi^{'}.\Pi^{'}}}{f_{\pi}}\right)
\end{equation}
The action may now be written
\begin{equation}
I=\frac{f^{2}_{\pi}}{2}\int d^{8}z trGG^{\dagger}
\label{L1}
\end{equation}
where we have introduced the matrix (dropping the primes)
\begin{equation}
G=\exp (i\tau . \Pi /f_{\pi})
\end{equation}
to display explicitly the previous
statement that the pions are coordinates of the manifold of SL(2C).
We therefore consider the matrix valued chiral superfield
\begin{equation}
G(x,\theta,\overline{\theta})
 =
\exp\left(\frac{i}{f_{\pi}}\tau.\Pi\right)=g(y)+\sqrt{2}\theta\psi(y)+
\theta^{2}F_{G}(y)
\end{equation}
with component fields $g,\psi$ and $F_{G}$ given by
\begin{equation}
 g(x)=G(x,\theta,\overline{\theta})|, \ \ \
 \psi_\alpha(x)=\frac{1}{\sqrt{2}}D_\alpha
G(x,\theta,\overline{\theta})|,
\ \ \
 F_G(x)=-\frac{1}{4} DDG(x,\theta,\overline{\theta})| \end{equation}
where $D_\alpha$ is the supersymmetric covariant derivative and $|$
indicates that the expression is evaluated at
$\theta=\overline{\theta}=0$.

We obtain
\begin{equation}
g=g(\pi(x))=\exp\left(\frac{i}{f_{\pi}}\tau.\pi\right)
\end{equation}
\begin{equation}
\psi_{\alpha}=\lambda^{a}_{\alpha}\frac{\partial
g}{\partial\pi^{a}}
\label{vielbein} \end{equation}
and
\begin{equation}
F_{G}=-\frac{1}{2}\lambda^{b}\lambda^{a}
\frac{\partial}{\partial\pi^{b}}\left(\frac{\partial
g}{\partial\pi^{a}}\right)
+F^{a}\frac{\partial g}{\partial \pi^{a}}
\label{FGbits}
\end{equation}
These closed expressions facilitate the geometrical
interpretation of our model
namely that we can relate the fermions $\psi$ to be fermions
defined in the
tangent space to the K\"ahler manifold at the point
($\pi^a,\pibar^{a}$),
related to $\lambda^a$ by the vielbein, $e^A_b(\pi)$, given by
$$ e^A_b(\pi) \ =
  \ \frac{1}{2}tr\left( \tau^A \frac{\partial
g}{\partial \pi^{b}} \right) $$
so that eq.(\ref{vielbein}) may be rewritten
\begin{equation}
\psi_{\alpha}=   \delta_{AB} \tau^A e^B_c(\pi)
\lambda^{c}_{\alpha}
 \end{equation}

Because supersymmetry requires that the scalar particles (the
pions)
be complex, contact
with the bosonic chiral Lagrangian can only be made below the
supersymmetry breaking
scale, $\mu$, at which the imaginary parts of the pions,
$\pi^a_I$,
acquire a mass. Although this scale is assumed to be far below
the
chiral symmetry breaking scale, $f_\pi$, it would be unnatural
to
include a term in the action which did not respect the initial
SU(2)$_L \otimes$SU(2)$_R$ symmetry. Thus we propose a soft
supersymmetry
breaking term
\begin{equation}
I_{SOFT} \ = \ - \int d^8z \frac{\mu^2}{4} \theta^2
\overline{\theta}^2
    \left(  tr \left( H \overline{H} \right)-2 f_\pi^2 \right)
\end{equation}
In terms of the pions (after substituting for the $\sigma$
fields,
 eq.(\ref{vev})) this
gives a mass term for the $\pi^a_I$ and a sequence of
interaction terms,
 between the real ($\pi^a_R$) parts and the imaginary parts
of the pions.
\begin{equation}
I_{SOFT} \ = \ \int d^4x \left( - \frac{\mu^2}{2} \pi^{a \
2}_I
   - \frac{\mu^2}{16 f_\pi^2}  \left( \pi^{a}_R \pi^{a}_I
\right)^2
   +  ....  \right)  \end{equation}
Thus we see that the soft supersymmetry breaking term provides
a mass
for the imaginary part of the pion, leaving the real part
(which we interpret as the usual pion)  massless. The
interactions
will undoubtedly have an effect, but they are all suppressed
by powers
of $\mu^2/f_\pi^2$ and thus expected to be small. We note here
that
the soft breaking term does not directly contribute a mass
to the fermions, $\lambda^a$. However as their masslessness is
no longer protected by supersymmetry, they would be expected
to
acquire a mass through higher order interactions. This mass is
expected to be of the order of the supersymmetry breaking
scale,
$\mu$, but it is quite possible that the fermion masses will
be
somewhat smaller than $\mu$, and so the effect of fermion
loops
will persist some way below the supersymmetry breaking scale.

We now consider the extension of the model to incorporate terms higher
than first order in momentum - the rule is simple; we must include all
terms consistent with the SU(2)$_{L}\otimes$SU(2)$_{R}$ symmetry and
supersymmetry, whilst recovering the usual bosonic chiral
 Lagrangian in the appropriate limit.
Manifestly, any term containing any power of the K\"ahler
potential, $V=trGG^{\dagger}$, satisfies the first constraint.
Since $V$ is a vector
superfield we immediately look at the standard kinetic term for a
vector
supermultiplet which by construction is of fourth power in momentum.

\noindent Introducing the matrix valued chiral superfield
\begin{equation}
W_{\alpha}=-\frac{1}{4}\overline{D}^{2}D_{\alpha}GG^{\dagger}
\label{Walpha}
\end{equation}
we include in our Lagrangian the manifestly supersymmetric term
\begin{equation}
\frac{1}{4}D^{2}(trW^{\alpha})(trW_{\alpha})
=\frac{1}{2}(trD)^{2}-\frac{1}{4}trF^{lm}trF_{lm}
-itr\eta.\sigma^{m}\partial_{m}tr\overline{\eta}
\label{nextbit}
\end{equation}
where in components we have
\begin{eqnarray}
\eta_{\alpha} &=& \sqrt{2}\psi_{\alpha}\Fbar_{G}
   -i\sqrt{2}\sigma^{m}_{\alpha\alphadot}
(\partial_{m}g)\psibar^{\alphadot}\nonumber\\
\frac{1}{2}D &=& F_{G}\Fbar_{G}-\partial_{m}g\partial^{m}g
+\frac{i}{2}(\partial_{m}\psi )\sigma^{m}\psibar
-\frac{i}{2}\psi\sigma^{m}\partial_{m}\psibar
\end{eqnarray}
and, as its name suggests, $F_{lm}$ is the 4-dimensional curl of the
vector
field component, $v_{m}$, of the vector superfield $GG^{\dagger}$,
where
\begin{equation}
v_{m}=\psi\sigma_{m}\psibar + i(g\partial_{m}\gbar -
\partial_{m}g.\gbar)
\end{equation}
Upon expansion in terms of component fields eq.(\ref{nextbit})
rapidly
becomes cumbersome, but it is supersymmetric and invariant under
SU(2)$_{L}\otimes$SU(2)$_{R}$ and it only remains for us to
demonstrate that
the usual bosonic chiral Lagrangian is recovered in the limit where
fermionic
fields are suppressed and the scalar fields are taken to be real.
Recalling eqs(\ref{FGbits} and \ref{Fterms}) we see that $F_{G}$ is
quadratic in fermionic fields to leading order and so in this limit
$tr\eta$ vanishes and $trD$ is simply
\begin{equation}
tr\left( \partial_{\mu}U\partial^{\mu}U^{\dagger} \right)
\end{equation}
where $U$ is the unitary matrix $\exp (i\tau . \pi /f_{\pi} )$ with
$\pi$ now a real scalar field. $(TrD)^{2}$ therefore reduces to
\begin{equation}
tr\left( \partial_{\mu}U\partial^{\mu}U^{\dagger} \right)
tr\left( \partial_{\nu}U\partial^{\nu}U^{\dagger} \right)
\end{equation}
which is one of the second order terms in the usual bosonic chiral
Lagrangian.

Similarly, $v_{m}$ and $trF_{lm}$ are now given by
\begin{eqnarray}
v_{m} &=&
U^{\dagger}\partial_{m}U\h-\h\partial_{m}U^{\dagger}.U\h=\h2U^{-
1}\partial_{m}U
\nonumber\\
trF_{lm} &=& 2tr \left[ U^{-1}\partial_{m}U.U^{-1}\partial_{l}U
                       -U^{-1}\partial_{l}U.U^{-1}\partial_{m}U
\right]
\end{eqnarray}
where we have exploited the fact that $U$ is a unitary matrix.
$TrF_{lm}$
vanishes (as expected from the fact that $v_{m}$ is a pure gauge) and
so
eq.(\ref{nextbit}) only contributes one term to the higher order
parts
of the bosonic chiral Lagrangian.

Returning to eq.(\ref{Walpha}) we are able to form a second,
independent,
SU(2)$_{L} \otimes$SU(2)$_{R}$ invariant
\begin{equation}
\frac{1}{4}D^{2}tr\left( W^{\alpha}W_{\alpha} \right)
\end{equation}
This time we need to consider $trD^{2}$ and $trF^{lm}F_{lm}$. $Tr
D^{2}$
gives us nothing new, but, whereas $trF_{lm}$ vanishes in the bosonic
limit,
$trF^{lm}F_{lm}$ does not and we recover the term
\begin{equation}
tr \left( \partial_{l}U^{\dagger}\partial_{m}U \right)
tr \left( \partial^{l}U^{\dagger}\partial^{m}U \right)
\end{equation}

\noindent This is the second and final addition to the
first order term required
 to duplicate the bosonic chiral Lagrangian. From this point of view
therefore,
the full supersymmetric Lagrangian to second order can be written
\begin{eqnarray}
{\cal L}&=&D^{2}\overline{D}^{2}trGG^{\dagger}\h +\h
  \alpha D^{2}(tr W^{\rho})(trW_{\rho})\h +\h \beta
D^{2}tr(W^{\rho}W_{\rho})
\end{eqnarray}
where $\alpha$ and $\beta$ are arbitrary coefficients corresponding
to the
two arbitrary coefficients in the higher order terms of the bosonic
chiral
Lagrangian.

So far we have been considering terms which are by construction
invariant
under the supergauge transformation
\begin{equation}
V\rightarrow V+\Phi +\Phibar
\end{equation}
where $\Phi$ is an arbitrary chiral superfield. There is no a priori
reason
to impose this constraint and we now go on to consider terms which do
not
obey this symmetry (whereas these terms are not expected to give us
anything new in the bosonic limit, they are allowed by supersymmetry
and
will be necessary for higher order supersymmetry effects).

We introduce the manifestly chiral (matrix valued) superfield
\begin{equation}
Z\h=\h -\frac{1}{4}\overline{D}^{2}GG^{\dagger}\h=\h
Z_{\phi}+\sqrt{2}\theta Z_{\lambda}+\theta^{2}Z_{F}
\end{equation}
with component fields
\begin{eqnarray}
Z_{\phi} &=& g\Fbar_{G} \nonumber \\
Z_{\rho\lambda} &=& \psi_{\rho}\Fbar_{G} +
ig\sigma^{m}_{\rho\alphadot}\partial_{m}\psibar^{\alphadot}\nonumber
\\
Z_{F} &=& F_{G}\Fbar_{G}+g\Box\gbar -
i\psi\sigma^{m}\partial_{m}\psibar
\end{eqnarray}
where we note that
$Z_{F}$ is the only component field containing a term which
does not depend explicitly on the fermions.

The first order term involving Z,
\begin{equation}
D^{2}trZ=D^{2}\overline{D}^{2}trGG^{\dagger}
\end{equation}
simply repeats eq.(\ref{L1}). Proceeding,
we form the next order manifestly supersymmetric term
\begin{equation}
D^{2}(ZZ)=2Z_{\phi}Z_{F}-Z_{\lambda}Z_{\lambda}
\label{p3bit}
\end{equation}
noting that each term in this expression is at least quadratic in
fermions.
Most importantly therefore, this term vanishes in the bosonic limit.

A typical term in eq.(\ref{p3bit}) (from $Z_{\lambda}Z_{\lambda}$) is
\begin{equation}
\psi\psi\Fbar_{G}\Fbar_{G}
\end{equation}

\noindent and using our earlier observations that
$F_{G}$ is quadratic in fermions to leading order and that
fermions have a factor $\sqrt{p}$
associated with them we see that this term is of order $p^{3}$. We
therefore
have a completely novel feature belonging to the supersymmetric form
of the
chiral Lagrangian, namely the next term in the momentum expansion is
of
order $p^{3}$ - filling the gap in the usual bosonic model.

Proceeding to fourth order in momentum we have the additional terms
\begin{equation}
D^{2}\overline{D}^{2}(Z\overline{Z})
\label{p4xtra1}
\end{equation}
and
\begin{equation}
D^{2}(ZZZ)=
3\left(Z^{2}_{\phi}Z_{F}-Z_{\lambda}Z_{\lambda}Z_{\phi}\right)
\label{p4xtra2}
\end{equation}
where we note again that eq.(\ref{p4xtra2}) is at least quartic in
fermionic fields and so vanishes in the bosonic limit. The term
(\ref{p4xtra1})
only differs from the term $D^{2}(W^{\alpha}W_{\alpha})$
already considered by the addition of
$$\frac{i}{4}D^{2}\overline{D}^{2}\sigma^{m}_{\alpha\alphadot}
(D^{\alpha}
GG^{\dagger})(\overline{D}^{\alphadot}\partial_{m}GG^{\dagger})$$
This latter expression vanishes when fermionic fields are suppressed
and so
the effect of (\ref{p4xtra1}) is also to add to the
effective chiral Lagrangian new terms which are only present in the
supersymmetric extension.

To obtain meaningful terms in the Lagrangian from the expressions in
eqs(\ref{p3bit}, \ref{p4xtra1}, \ref{p4xtra2}) we need to take
the trace in every possible
independent way. We see that there will be two terms from
eq.(\ref{p3bit}),
two from eq.(\ref{p4xtra1}) and a further three from
eq.(\ref{p4xtra2}). Each
will have arbitrary coefficients and, where appropriate, the
hermitian
conjuate term is to be added, again with an arbitrary coefficient.

We thus see that the number (14) of independent next to leading order
terms,
and hence the number of arbitrary coefficients to that order, is much
larger
than for the bosonic case (where there are just two).
This leads to a much richer structure for the
effective action, despite the fact that all but
two of these terms become unimportant
 below the supersymmetry breaking scale, $\mu$.

It is easy to see how this formalism can be extended to
higher chiral symmetry,\\ SU(N)$_L \otimes$SU(N)$_R$. This is
simply achieved by defining the matrix $H$ of superfields to
be an $N \times N$ matrix, transforming as an ($N,
\overline{N}$) of
the chiral symmetry. The term $det(H) \Phi$ in the action,
eq.(\ref{action1}),
now generates higher order, nonrenormalisable terms. This does
not bother us
since the linear version is just taken as a guide to construct
the
effective chiral action. The constraint on the $\sigma$
fields, eq.(\ref{vev}),
now becomes an $N^{th}$ order equation, from which the metric
on the
K\"ahler SL(NC) manifold can in principle be determined,
although
the algebra now becomes intractable.  \bigskip

\newpage
\noindent {\bf Acknowledgements}\\
We are grateful to Terry Elliott, Steve King
and Peter White for useful conversations, particularly with reference
to
the construction of the superpotential.
This work is partly supported by SERC grant no GR/J21569. \bigskip

During the preparation of this manuscript our attention was
drawn to
a paper by Clark and ter Veldhuis \cite{clark}, who by
approaching
the subject from a different viewpoint, namely by taking the
limit of
the minimal supersymmetric standard model as the Higgs mass
becomes infinite,
have arrived at a model which coincides with the leading
momentum
term of the model presented here. 

\end{document}